\documentclass[]{tADP2eMOD}

\usepackage{epstopdf}
\usepackage{subfigure}
\usepackage{lmodern}

\theoremstyle{plain}

\theoremstyle{definition}

\theoremstyle{remark}

\newcommand{\figwid}{0.47\textwidth}

\begin{document}



\title{Statistical and computational intelligence approach to analytic continuation in Quantum Monte Carlo}

\author{
\name{Gianluca Bertaina\textsuperscript{a}, Davide Emilio Galli\textsuperscript{a}$^{\ast}$\thanks{$^\ast$Corresponding author. Email: davide.galli@unimi.it} and Ettore Vitali\textsuperscript{b}}
\affil{\textsuperscript{a}Dipartimento di Fisica, Universit\`a degli Studi di Milano, via Celoria, 16 - 20133 Milano, Italy;
\textsuperscript{b}Department of Physics, College of William and Mary, Williamsburg, Virginia 23187-8795, USA}
}

\maketitle

\begin{abstract}
The term analytic continuation emerges in many branches of Mathematics, Physics, and, more
generally, applied Science. Generally speaking, in many situations, given some amount of
information that could arise from experimental or numerical measurements, one is interested
in extending the domain of such information, to infer the values of some variables
which are central for the study of a given problem. For example, focusing on Condensed
Matter Physics, state-of-the-art methodologies to study strongly correlated quantum physical systems
are able to yield accurate estimations of dynamical correlations in imaginary time. Those functions
have to be extended to the whole complex plane, via analytic continuation, in order to infer
real-time properties of those physical systems. In this Review, we will present the Genetic Inversion via Falsification
of Theories method, which allowed us to compute dynamical properties of
strongly interacting quantum many--body systems with very high accuracy.
Even though the method arose in the realm of Condensed Matter Physics,
it provides a very general framework to face analytic continuation problems
that could emerge in several areas of applied Science.
Here we provide a pedagogical review that elucidates the approach we have developed.
\end{abstract}

\begin{classcode}
02.30.Zz Inverse problems, 67.10.Fj Quantum statistical theory, 67.25.dt Sound and excitations,
02.70.Ss Quantum Monte Carlo methods
\end{classcode}

\begin{keywords}
analytic continuation; inverse problems; genetic algorithms; quantum liquids; dynamics; elementary excitations
\end{keywords}

\section{Introduction}
A very challenging problem emerging in pure and applied Physics, as well
as in many branches of Science, is analytic continuation.
Such term arises naturally, at an abstract level, in the realm of complex
analysis, where it is defined as a technique to extend the domain
of a given analytic function, say $F:\Omega\subset \mathbb{C} \to \mathbb{C}$. 
Strictly speaking, performing analytic continuation means finding an analytic function $\tilde{F}:\tilde{\Omega} \subset \mathbb{C} \to \mathbb{C}$ such that $\tilde{\Omega} \supset \Omega$ and
$\tilde{F}(z) = F(z)$ $\forall z \in \Omega$.
In other words, the key point is to use the information encoded in $F(\Omega)$
to find, or infer, the values of $F$ on a wider set.
In addition, in many important cases, the knowledge of $F(\Omega)$
itself can be affected by uncertainties, that could arise from the
numerical or experimental determination of the values of the function.

Situations where analytic continuation turns out to be useful or
even necessary are present in a very
broad range of physical or even more
generally scientific studies, encompassing Quantum Field Theory,
Condensed Matter Physics, as well as image reconstruction and many others. 
 
A wide family of physical applications of analytic continuations
originate from the celebrated {\it{Wick rotation}}, a mapping
between real time and imaginary time:

\begin{equation}
\label{an-cont}
F(t) \longleftrightarrow F(-i \tau ) \quad ,
\end{equation}
whose importance and usefulness is essentially
due to mathematical reasons.
For example, in the realm of Quantum Field Theory, the Euclidean
space--time approach provides much more well-behaved
and well-defined expressions than the formulation in Minkowski
space--time \cite{zinn2002quantum}. The relation between the two approaches is an
analytic continuation problem.

Another central example, which will be the topic of this
review, arises in Condensed Matter physics.
In this context, the Wick rotation provides a mapping between the 
quantum mechanical evolution operator and the imaginary-time propagator,
or thermal density matrix:

\begin{equation}
\label{an-cont-prop}
e^{-\frac{i t \hat{H}}{\hbar}} \longleftrightarrow e^{-\frac{\tau \hat{H}}{\hbar}}\quad ,
\end{equation} 
where $\hat{H}$ is the Hamiltonian operator of a quantum system and $\hbar$ is Planck's constant.
As in Quantum Field Theory, calculations involving the imaginary-time propagator are generally more well-behaved, and there exist
extremely accurate techniques to compute imaginary-time correlation
functions. In particular, most 
quantum Monte Carlo (QMC) methodologies, which
nowadays are crucial for the study of strongly correlated physical 
systems \cite{hammond1994monte,gubernatis2016quantum}, are intrinsically formulated in imaginary time,
and yield estimations of correlation functions involving
the imaginary-time propagator in Eq.~\eqref{an-cont-prop}.
It is thus necessary and, as we will discuss below, very challenging,
to perform the analytic continuation necessary to infer real-time properties.

Incidentally, we mention another context in which analytic continuation turns out to
be extremely useful: the reconstruction of images. Consider, for example, the phase retrieval problem in Coherent Diffractive Imaging (CDI) \cite{miao_extending_1999}. This technique uses the measured diffraction pattern of a coherent beam scattered by an object to obtain spatial information. In an experimental far-field intensity measurement, the diffracted intensity, for weak scattering objects, is proportional to the modulus of the Fourier Transform of the object scattering function. Any information on the phase is lost in the far-field measurement, and has to be retrieved in order to obtain the scattering function; this is usually attained by means of suitable algorithms. The problem of reconstructing the full information (modulus and phase) in Fourier space from the limited set of data (partial and noisy measurement of the modulus) is thus a very important example of analytic continuation problem.

\subsection{Analytic continuation and inverse problems}
Before digging further in this field, we find very important to
mention that analytic continuation can be embedded in
a much wider family of problems, that are called {\it inverse problems}.
In this very general context, the relation between {\it theories}
and {\it observations} is the central point.
Given a theory, we can of course, at least in principle, predict the results of
the observations, this being the {\it{direct problem}}, but the
{\it{inverse problem}}, namely to deduce a theory
from observations, is naturally much more subtle. 
The key question, in a schematic way, is the following:
when building up a {\it theory}, how many answers may
we expect from {\it observations}?

Even if, at first sight, this seems to have nothing to do with analytic continuation,
a connection can be foreseen if one considers the imaginary-time
properties as the {\it observations} and the real-time
properties as the {\it theory}. This interpretation, although
somehow artificial, is indeed meaningful in the sense
that, in fact, techniques are available to compute, i.e.
to {\it{observe}}, imaginary-time properties, while, currently,
general direct computation in real time is much harder. Therefore,
real-time properties have to be guessed, like a theory
is guessed from observations. Anyway, inverse problems have been well 
known since the earliest days of research in Physics
and the relation between theory and experiment is, of course,
central in Physics and in Science in general, much
beyond the domain of analytic continuation.

The key question quoted above, in the more specific language
of analytic continuation, sounds as follows:
how can we use the observations of the values of a function $F$ on a given domain, say
imaginary time, to infer the function $\tilde{F}$ on a different domain, real
time, where we cannot calculate the values directly? At a first glance, one feels
that such an inverse procedure in realistic situations is
unavoidably ill--posed, since any set of observations is
limited and noisy, thus ruling out the possibility of
finding out one and only one theory, i.e. function, whose predictions fit
such data.
In other words, there will exist infinite functions $\tilde{F}$
defined inside the complex plane that,
when restricted to imaginary-time axis, will be compatible
with the estimated imaginary-time correlation functions.

Many tools have been devised to face those ill--posed inverse problems.
Roughly speaking, the approaches can be divided into two highly overlapping
families. Some approaches modify the problem, via a regularization technique,
attempting to find a well-posed, or less ill-posed, problem. 
Regularizing the problem means introducing additional constraints
to the problem itself; such constraints very often rely on euclidean norms 
and are meant to have a unique solution to the problem or, at least, to drastically 
reduce the number of possible solutions \cite{kaipio_statistical_2005}.
On the other hand, another class of approaches looks for a solution in a statistical 
meaning, aiming at maximizing the probability of finding a function $\tilde{F}$, given
the function $F$ which is known \cite{kaipio_statistical_2005}. This probability is built up in the
realm of a Bayesian description of the analytic continuation problem.
In many situations, mixed approaches including regularization techniques
and statistical approaches are used.

Our approach relies on a {\it{falsification principle}} which we are now going do explain in some detail.
Following Popper \cite{popper2002logic}, Tarantola \cite{tarantola_popper_2006} put forward the proposition that
observations may only {\it falsify} a theory.
In order to understand the far reaching consequences of this
proposition, let's formulate the problem at an abstract level.
Suppose we have measured $F(\Omega)$, through some experimental
or numerical tool. In principle, if we were able to perform an infinite
number of measurements, we would have found a sample of results.
It is thus natural to consider our data, which we will
denote by $d$, as a point in a, possibly huge, set of all possible outcomes,
that is a set of data, say $\mathcal{D}$.
On the other hand, we denote $\mathcal{S}$ the set of all the possible theories,
i.e. the set of all the candidate solutions. The inverse problem would sound as follows:
may we find a theory $s \in \mathcal{S}$ predicting $\mathcal{D}$?
That is, can we find $s \in \mathcal{S}$ which is {\it not falsified}
by any of the elements of $\mathcal{D}$?
If the answer was positive, of course, that theory would be
our solution.
This is an idealized situation, since in a finite time we actually never know
$\mathcal{D}$ but only a subset $\mathcal{D}^{\star} \subset \mathcal{D}$
whose elements are in general used
to give an estimation of the statistical uncertainties in the observations.
Normally an infinite number of theories exists
compatible with $\mathcal{D}^{\star}$; it is then clear that we may exclude some theories
but we still remain with a set $\mathcal{S}_{\mathcal{D}^{\star}} \subset \mathcal{S}$
of equivalent ``solutions''.

Depending on the mathematical details of the space $\mathcal{S}$,
a natural idea appears to be that of devising a procedure
enabling to capture what the theories in $\mathcal{S}_{\mathcal{D}^{\star}}$
do have in common. In this way, even if we won't succeed in finding
out a unique theory $s \in \mathcal{S}$, we will be able nevertheless
to find out a class of features, providing physical properties, that $s$ has to possess
so that it will not be falsified by the limited set of observations.

In order to better understand what this means in a specific
example, we will now focus on the typical analytic continuation
problem in condensed matter physics: the estimation of spectral functions of 
many--body quantum systems starting from imaginary-time correlation functions.

\section{Analytic continuation in Quantum Monte Carlo}
The formal definition of a spectral function of a many body physical system is as follows:

\begin{equation}
\label{spectral}
s(\omega) = \int_{-\infty}^{+\infty}\frac{dt}{2\pi}e^{i\omega t}\langle e^{\frac{i\hat{H} t}{\hbar}}
\hat{A}e^{-\frac{i\hat{H} t}{\hbar}}\hat{B}\rangle \quad ,
\end{equation}
$\hat{A}$ and $\hat{B}$ being given operators acting on the Hilbert space of the system
whose Hamiltonian operator is $\hat{H}$. The brackets indicate the expectation value on
the ground state or thermal average. We also introduce the imaginary-time correlation function:

\begin{equation}
\label{imaginary}
F(\tau) = \langle e^{\frac{\hat{H} \tau}{\hbar}} \hat{A}e^{-\frac{\hat{H} \tau}{\hbar}}\hat{B}\rangle\quad.
\end{equation}
It is evident that, if we were able to extend the domain of $F$,
computing:

\begin{equation}
\label{complex}
\tilde{F}(z) = \langle e^{\frac{i \hat{H} z}{\hbar}} \hat{A}e^{-\frac{i \hat{H} z}{\hbar}}\hat{B}\rangle, \quad z \in \mathbb{C}\quad,
\end{equation}
we would immediately be able to compute Eq.~\eqref{spectral}.

In the language of inverse problems, the function $s(\omega)$ is
the {\it{theory}} we are looking for, while $F(\tau)$ corresponds
to the {\it{observations}}. In the language of analytic continuation, on the
other hand, it is evident that the two functions are related by Eq.~\eqref{an-cont-prop} and a Fourier transform, or, equivalently, a single inverse Laplace transform. As mentioned earlier, the reason why we consider $F(\tau)$ as 
the observation, while $s(\omega)$ is the unknown theory, is that,
using quantum Monte Carlo simulations, when the sign problem
does not show up, for example for Bose fluids, it is possible to
obtain exact estimations of the values of the function $F$ \cite{boninsegni_density_1996,baroni_reptation_1999,vitali_initio_2010}.
In this context, exact means that, for a given statistical
accuracy, every systematic error can be reduced below the noise
level via a suitable tuning of the parameters.

To summarize, the situation is as follows: with QMC methods
we can estimate values of $F(\tau)$ in correspondence
with a finite number of imaginary-time values depending on the discretization of the methodology. To be specific we will use the
notation $f\equiv\{f_i=F(i \delta \tau),0\le i < l\}$. 
In general $f$ is obtained as an average of several
QMC calculations of $F(\tau)$, each affected by statistical noise, and which are used to estimate the
{\it statistical uncertainties} $\{\sigma_i\}$ associated with $\{f_i\}$.
Moreover, the set of observations can be often enriched relying
on sum rules, which prompt to perform additional QMC measurements
providing estimations for some momenta of
$s(\omega)$: $c \equiv \{c_n=\int_{-\infty}^{+\infty}
d\omega \omega^n s(\omega),\,n\in\mathbb{Z}\}$ (for example $c_0=\langle \hat{A}\hat{B}\rangle$
may be easily estimated in equilibrium QMC simulations with an associated statistical uncertainty). 

In this context, the inverse problem of estimating $s(\omega)$ has
the formal appearance of a Fredholm integral equation
\begin{equation}
\label{problem}
F(\tau) = \int_{-\infty}^{+\infty}d\omega\mathcal{K}(\tau,\omega)s(\omega)\quad,
\end{equation}
where for example, at zero temperature,
$\mathcal{K}(\tau,\omega) = \theta(\omega)e^{-\tau\omega}$, $\theta(\omega)$
being the Heaviside distribution.
In many cases, this equation can be complemented
by some {\it a priori knowledge} that can be deduced from
the formalism of quantum mechanics, such as the support,
non--negativity or some further properties.

Since we start from limited and noisy data and the considered kernel corresponds to an ill-conditioned matrix when discretized,
it is evident that Eq.~\eqref{problem} does not have a unique solution and that small variations of the data can largely affect the resulting spectra.
In general, there will be an infinite number of functions $s(\omega)$ whose ``predictions'', namely the reconstructed data $\bar{f}$ obtained by the
integral in Eq.~\eqref{problem}, fall inside
the confidence intervals $(f_i - \sigma_i,f_i + \sigma_i)$.

\section{Genetic Inversion via Falsification of Theories}\label{sec:gift}
Being of paramount importance, the task of facing the problem in Eq.~\eqref{problem} has been investigated by many methods.

The Maximum Entropy Method (MaxEnt) and its variants \cite{bryan_maximum_1990,silver_maximumentropy_1990,jarrell_bayesian_1996,gunnarsson_analytical_2010,burnier_bayesian_2013} are the most popular approach. MaxEnt applies the maximum likelihood principle within a Bayesian framework. To be more specific, the Bayesian conditional probability of the spectrum $s$, given the measurements $f^*$, is equivalent to $P(s|f^*)= P(f^*|s)P(s)/P(f^*)$, where the conditional probability $P(f^*|s)$ of the measurements, given the spectrum, is called the \emph{likelihood}, while the \emph{a priori} probability $P(s)$ reflects previous belief concerning the spectrum in the absence of data. Finally $P(f^*)$ is the normalization. In the classical formulation MaxEnt assumes $P(f^*|s)\propto \exp{(-\chi^2/2)}$, where $\chi^2$ is the quadratic distance of the data $f^*$ from the reconstructed data $\bar{f}$. The \emph{a priori} knowledge is enforced as an entropic term with respect to a default model $m$: $P(s)\propto \exp{(-\alpha \int d\omega s(\omega) \ln{[s(\omega)/m(\omega)]}})$, and various ways are proposed to treat the parameter $\alpha$ and the default model $m(\omega)$. The entropic term and some suitable regularization of the necessary matrix operations are crucial in this approach in order to obtain a smooth solution.

The Average spectrum method (ASM), also called Stochastic Analytical Inference  \cite{white_average_1991,sandvik_stochastic_1998,beach_identifying_2004,syljuaasen_using_2008,reichman_analytic_2009,fuchs_analytic_2010,sandvik_constrained_2015}, drops out the need of the entropic prior by averaging over different spectra according to the likelihood  $P(f^*|s)\propto \exp{(-\chi^2/{2T})}$. Through a Monte Carlo sampling of spectral functions, the effective temperature $T$ is used to gradually force more adherence to the data via a simulated annealing procedure. Various prescriptions for stopping the simulated annealing have been investigated. The averaging procedure smooths the final average and retains only common features. It has been demonstrated \cite{beach_identifying_2004} that MaxEnt can be derived as a mean-field limit of the ASM. This finding has been further explored in \cite{fuchs_analytic_2010}.

The Stochastic optimization with consistent constraints method (SOCC) \cite{mishchenko_diagrammatic_2000,goulko_numerical_2016,bao_fast_2016} also averages over spectra which are obtained with a Monte Carlo walk aiming at minimizing the $\chi^2$ distance of the data and reconstructed data; however, the averaging procedure consists of a linear combination of spectra, aiming at obtaining a maximally smooth result. This flexible averaging procedure allows for estimating error-bars for the spectrum, by artificially emphasizing the main spectral features up to values which would worsen the value of the $\chi^2$.   

Without being exhaustive, we mention that other research directions involve the use of other basis functions for the spectra \cite{beach_reliable_2000,schott_analytic_2016}, the exploit of more information from tailored QMC simulations \cite{dirks_extracting_2013,rota_quantum_2015} or the consideration of different kernels \cite{roggero_dynamical_2013}. 
We also note that, in the context of image reconstruction, different measures of distance of the reconstructed data from measurements have been proposed, such as the Kullback-Leibler or I-divergence \cite{csiszar_why_1991}, which allow for the implementation of deterministic error-reduction algorithms \cite{snyder_deblurring_1992}.

The GIFT approach \cite{vitali_initio_2010}, which we are going to describe in detail, follows more radically the general scheme outlined in the introduction:
we need a space of models $\mathcal{S}$, containing a wide collection
of spectral functions consistent with any {\it prior knowledge} about $s(\omega)$,
a falsification procedure relying on the QMC ``measurements'' $d=\{f,c\}\in \mathcal{D}$, and
a strategy to capture the accessible physical properties of $s(\omega)$. The introduction of the space $\mathcal{D}$ is meant
to stress that there is nothing special in the
particular set of QMC measurements $d$. If a new independent
simulation is performed, a new set of measurements
will show up and it will be completely equivalent to the
original one.

\subsection{The space of models}\label{subsec:frequency}
In our mathematical framework $\mathcal{S}$ is made of step functions, providing
a compromise between the possibility of suitably approximating
{\it any} model of spectral function and the feasibility of numerical operations
inside it. 

In the typical case ($\hat{A}=\hat{B}^{\dagger}$) when $s(\omega)$ is known to be real-valued,
non-negative and the zero-momentum sum rule holds, we rely on models $\overline{s}$ of the form:
\begin{equation}
\label{solutions2}
\overline{s}(\omega) = \sum_{j=0}^{m-1}\frac{s_j}{\mathcal{M}\Delta\omega_j}\chi_{I_j}(\omega),
\quad \sum_{j=0}^{m-1}s_j = \mathcal{M} \quad .
\end{equation}

We rely on a fixed partition $\{\omega_0,...,\omega_m\}$ of widths $\Delta\omega_j$ of an interval of the real line much larger than the hypothesized support of $s(\omega)$. In particular, we use equally-spaced frequencies up to an intermediate value where the spectrum is hypothesized (or verified with exploratory reconstructions) to have significantly decayed, and we then employ exponentially spaced frequencies, to ease the fulfillment of high-order sum rules. In some cases, for example for some 1D models \cite{sandvik_constrained_2015,bertaina_onedimensional_2016,motta_dynamical_2016}, a minimal threshold frequency $\omega_{th}$ is known, below which the spectrum is zero; this can be easily implemented by setting $\omega_0=\omega_{th}$.
We use the characteristic function $\chi_{I_j}(\omega)$ of the intervals $I_j = [\omega_j,\omega_{j+1})$, which takes the value $1$
inside $I_j$ and $0$ outside.
Moreover, we introduce a discretization of the codomain, $s_j \in \mathbb{N}\,\cup\{0\}$, to make the space finite.
$\mathcal{M}$ provides the maximum number of quanta of spectral weight available
for the ensemble of the intervals $I_j$. Notice that $\overline{s}(\omega)$ differs from
the physical spectral functions by a factor $c_0$, being $c_0$ the zero--momentum,
which belongs to the set of observations.

Just to stress the different kind of applications that the present framework could include, we note that the choice expressed by Eq.~\eqref{solutions2} originates from the fact that the first GIFT implementation was inspired by an application in quantitative finance related to a stochastic optimization of portfolios based on Genetic Algorithms \cite{galli__2001}. In that application $m$ represented the maximum number of assets included in the portfolio, $M$ the total investment, which is naturally quantized, and $s_j$ was the number of quanta of investment for asset $j$.
In the present context, the choice in Eq.~\eqref{solutions2} is meant to provide a sufficiently vast family of functions: the basic idea is that, apart from
a priori knowledge arising from the formalism of quantum mechanics, no
additional information has to be imposed in the definition of the space of
model. Of course, one has to verify that the results do not change significantly when decreasing the interval widths. Other representations of the spectral function, such as a sum of delta contributions, may be implemented as well, with negligible impact on the methodology. Other stochastic approaches also optimize the widths of the intervals \cite{mishchenko_diagrammatic_2000} or the positions of the delta functions \cite{fuchs_analytic_2010}: this can increase efficiency.

\subsection{Falsification, Fitness, and Genetic Algorithms}
How can we explore $\mathcal{S}$ and falsify its elements? As mentioned before, the most important point is the translation into a practical algorithm
of the {\it{falsification principle}} described in the introduction.
In principle, not only the observed data $d = \{f,c\}$, but any
{\it equivalent data} $d^{\star}=\{f^{\star},c^{\star}\}$, that are the result of an independent simulation, should
play an equivalent role in determining whether a model has to be falsified or not.
The simplest way to achieve this in practice leads us to the definition of the
{\it fitness} of model $\bar{s}$:

\begin{equation}
 \Phi_{d^{\star}}(\overline{s}) =
-\sum_{j=0}^{l-1} \frac{1}{\sigma_j^2}\left[f_j^{\star} - c_0^{\star}\int d\omega \, e^{-\omega j \delta\tau}
\overline{s}(\omega)\right]^2 
- \sum_n \gamma_n \left[c_n^{\star} - c_0^{\star}\int d\omega \,\omega^n
\,\overline{s}(\omega)\right]^2
\label{fitness}
\end{equation}
depending on the set of data, together with the introduction of a scheme to build up
$d^{\star}$ starting from $d$. In principle, one could store different realizations $d^\star$ directly from independent blocks in the QMC simulations.
In our implementation, different random sets $d^{\star}=\{f^{\star},c^{\star}\}$ are
obtained by resampling independent Gaussian distributions centered on the original QMC observations $d$,
with variances which correspond to the estimated QMC statistical uncertainties. Generalizations can be easily conceived if the covariance matrix
among the data is computed during a simulation. More precisely, suppose
that the $l \times l$ matrix $\mathcal{C}_{ij} = cov(f_i, f_j)$
is estimated. The above-mentioned sampling of independent Gaussian distributions relies
on the approximation $\mathcal{C}_{ij} \simeq \delta_{ij} \sigma_i^2$. If the nondiagonal elements are also computed, the equivalent data can be sampled using
the formula:
\begin{equation}
f^{\star}_i = f_i + \sum_{j} L_{ij} \varepsilon_j \quad,
\end{equation}
where $\varepsilon_j $ are realizations of independent standard normal
random variables, and the lower-triangular matrix $L_{ij}$ satisfies
$LL^T = \mathcal{C}$ \cite{kaipio_statistical_2005}. Matrix $L$ can be obtained via standard Cholesky decomposition, since the covariance
matrix of vector $f$ is, by construction, real-valued and positive definite.
Once this is done, the fitness in Eq.~\eqref{fitness} can be modified
using the matrix $\mathcal{C}^{-1}=(L^{-1})^T L^{-1}$ instead of $\delta_{ij}/{\sigma_j^2}$. The results presented in this Review are obtained by ignoring covariance and 
considering only the variance of the data. It can be argued that the typical covariance of QMC imaginary-time data is positive, due to the kinetic term in the density matrix, so that ignoring covariance yields unnecessary fluctuations in our resampled imaginary-time data. This is overcome by pursuing lower values of the fitness function \eqref{fitness} before the final averaging procedure is done \cite{sandvik_stochastic_1998}.

In the definition of Eq.~\eqref{fitness}, the free parameters $\gamma_n > 0$ are adjusted in order to make the
contributions to $\Phi_{d^{\star}}$ coming from $f^{\star}$ and from
$c^{\star}$ of comparable order of magnitude, provided that convergence of the algorithm does not slow down due to a too strong constraint coming from high values of $\gamma_n$.
If it happens that one $c_n$ is exactly known, no error is added by making $c_n^{\star}=c_n$.

The idea, then, is as follows: we sample several independent equivalent data $d^{\star}$
and find out the set of models which are not falsified by them. 
In order to achieve this, we rely on Genetic algorithms (GA), which
are known to provide an extremely efficient tool 
to explore a sample space
by a nonlocal stochastic dynamics, via a survival--to--fitness evolutionary process
mimicking the natural selection we observe in the natural world.
Such evolution aims at maximizing the fitness towards ``good'' {\it building blocks} \cite{Goldberg_1989} which,
in our case, should recover information on physical spectral functions.

In our GA, for each resampled $d^{\star}$, we start randomly constructing a collection of $\overline{s}(\omega)$,
the {\it initial population}, consisting of $\mathcal{N}_{\overline{s}}$ individuals.
Each $\overline{s}(\omega)$ is coded by $m$ integers, $s_j$ in Eq.~\eqref{solutions2}.
The genetic dynamics then consists in a succession
of {\it generations} during which the initial {\it population},
consisting of $\mathcal{N}_{\overline{s}}$ {\it individuals}, is
replaced with new ones in order to reach regions of $\mathcal{S}$
where high values of the {\it fitness} exist, for a given $d^{\star}$.
In the passage between two generations a succession of ``biological--like'' processes takes place, given by the genetic operators described in the next subsection. The GA dynamics performs the falsification procedure: for each $d^{\star}$,
only the $\overline{s}(\omega)$ with the highest fitness in the last generation provides
a model for $s(\omega)$ which has not been falsified by $d^{\star}$. This yields the set $\mathcal{S}_{\mathcal{D}^{\star}}$ made of the
elements $c_0^{\star}\,\overline{s}(\omega)$.

Finally, an averaging procedure of the elements of $\mathcal{S}_{\mathcal{D}^{\star}}$
appears as the most natural way to extract physical information. Presently, we also calculate the variance of the ensemble $\mathcal{S}_{\mathcal{D}^{\star}}$ as a way to estimate the variance of the spectra. However, this is only qualitative, since a more complete information would stem from considering the whole \emph{covariance} matrix of the estimated spectral function. Notice also that the definition of the fitness in Eq.~\eqref{fitness} is essentially equivalent to the log-likelihood of other Bayesian approaches, in presence of a uniform prior. However we do not attempt to find a single maximum-likelihood spectrum, which would be affected by the saw-tooth instability, but we create an ensemble of spectra for which the magnitude of the fitness is of order of less than $l+1$. This is very similar to the ASM approach, with the advantage of the speed-up coming from nonlocal genetic evolution.

\subsection{The genetic dynamics}\label{subsec:dynamics}
The typical genetic operators are called {\it selection}, {\it crossover} and {\it mutation}. We found it useful to add also a {\it rejection} operator \cite{bertaina_onedimensional_2016}, that allows for more flexibility in the mutation moves, and constitutes a clean bridge between the genetic and the ASM approaches, yielding a hybrid genetic-ASM algorithm.
We now describe each step of the genetic evolution, for a given realization of $d^\star$. The {\it Selection} and {\it Rejection} operators are always executed, while the {\it Crossover} and various {\it Mutation} operators are called with some probability, which is chosen after performing small-scale exploratory runs to increase efficiency.
\begin{itemize}
 \item {\it Selection}. The population of the previous generation is ordered in ascending fitness; then a couple of individuals (``\emph{mom}'' and ``\emph{dad}'') are selected
corresponding to the indexes $k=[\mathcal{N}_{\overline{s}}\,\,$int$(r^{\beta})]+1$
obtained by sampling two uniform random numbers $r \in [0,1)$;
the nonlinearity of $k$ on $r$
is such that individuals with large fitness are preferentially selected; we typically use $\beta=1/3$. 
 \item {\it Crossover}. An amount of quanta $Q$ is uniformly chosen in the interval $\left[0,\mathcal{M}/2\right]$, to be exchanged between \emph{mom} and \emph{dad}. Optionally, one can choose $Q\approx\mathcal{M}/2$, which accelerates convergence, but is more computationally expensive. For both \emph{mom} and \emph{dad}, a random set of bins $\{I_j\}$ is chosen whose total spectral weight is $Q$. Then, the selected weights of the mum spectrum are moved to the \emph{dad} spectrum, keeping their original mum positions. Vice-versa, the selected \emph{dad} weights are moved to the mum spectrum, keeping their original \emph{dad} positions. This way, the total spectral weight is preserved, while a very nonlocal operation is performed. In particular, it is likely that the main spectral features of mum and \emph{dad} are exchanged. We observe that this operator, together with the selection operator, is the main feature concretely distinguishing our approach from the ASM and SOCC methods, allowing for a significant speed-up of the evolution. This operation is indeed performed with a high probability of $30\div40\%$.
 \item {\it Mutation}. Having obtained two new spectra, \emph{son}$_1$ and \emph{son}$_2$, from \emph{mom} and \emph{dad}, we are now prompted to perform single-spectrum moves. There is large room for experimentation at this point; however, since detailed balance is not respected, a risk is present to systematically bias the final result towards specific spectral features, depending on the chosen mutation operators. For example, in the relevant cases discussed in Subsection~\ref{subsec:3d}, typical spectra should include a single narrow peak and a minor broad structure, while in Subsection~\ref{subsec:1d} major broad structures are expected: in this case, mutation operators favoring only the identification of peaks may hamper the efficient reconstruction of almost flat spectra. It is thus important to include a variety of mutation operators to render the algorithm ergodic. We name a few examples:
 \begin{itemize}
  \item {\it Local mutation}. The shift of a random fraction of spectral weight between two random neighboring intervals. With small probability, the shift is performed preferentially to intervals where spectral weight is already present, which is useful for the quick discovery of peaks \cite{vitali_initio_2010}.
  \item {\it Nonlocal mutation}. The shift of a random fraction of spectral weight between two random intervals. In order to avoid the worsening of the fitness due to this nonlocal move, especially in the part concerning the spectral moments $c$, we impose a detailed balance condition using the probability density $\exp{(-\Phi_{d^{\star}}(\overline{s})/T)}$, as in the ASM method, with an effective temperature that is reduced during the simulation.
  \item {\it Smoothening}. A short succession of neighboring intervals is randomly chosen. The weights are convoluted with a smoothening kernel, which essentially performs a weighted average of the original spectral weights.
  \item {\it Error reduction}. The steepest descent method, or one of its more stable variants, is applied to the maximization of the fitness in Eq.~\eqref{fitness}, using its functional derivative with respect to the spectrum $\bar{s}$. Its efficient implementation suggests the use of real- instead of integer-valued weights, which requires a trivial adjustment of the algorithm. This operator performs a deterministic optimization of the spectra; however, when used alone, it is only able to get to local maxima of the fitness. The combined use of stochastic and deterministic operators yields a so-called \emph{memetic} algorithm \cite{Moscato_1992,ong_memetic_2010,MPR_article_2016}.
  \item {\it Flattening}. A random number of neighboring interval weights is substituted by their average. This is a very nonlocal operation and has to be selected not too often. It is in particular useful to explore almost flat and broad spectra, which would otherwise be created very seldom, due to entropic reasons.  
 \end{itemize} 
 After all the described operations, missing or exceeding weight is randomly redistributed among the spectral intervals, in order to respect the zero-momentum sum rule.
 \item {\it Rejection}. Since some of the mutations, such as \emph{flattening}, could bias the population, an accept/reject operation is carried out, by performing a Metropolis check comparing the weight $\exp{(-\Phi_{d^{\star}}(\overline{s})/T)}$ of \emph{son}$_1$ with \emph{mom} and of \emph{son}$_2$ with \emph{dad}. Notice that this does not implement detailed balance because of the \emph{selection} and \emph{crossover} operators. However, by disabling those, we essentially recover the ASM approach. The effective $T$ is typically the same used in the \emph{nonlocal mutation} move and is reduced through a simulated annealing schedule. We start from a very high $T\approx 10^{6}$, where the moves are usually accepted (analogously to the original version of the algorithm \cite{vitali_initio_2010}), and we geometrically reduce it every $\sim 100$ iterations, until a small value $T\approx 10^{-2}$ is reached. We observed that the precise initial and final $T$ have a negligible impact on the resulting final spectrum.
\end{itemize}

The above procedure is repeated until all the individuals $\overline{s}(\omega)$ of the population are replaced by a new generation,
except for the $\overline{s}(\omega)$ with the highest fitness in the old generation which
is cloned (\emph{elitism}).
To decrease the computational cost, the number of individuals in the new population is reduced by a small fraction at every generation
till $\mathcal{N}_{\overline{s}}$ is equal to a given minimal value; from this point over, the number
of individuals $\mathcal{N}_{\overline{s}}$ in the new generations is kept constant. We monitor the average fitness and $\chi^2$ of the best $\bar{s}(\omega)$ in the final ensemble $\mathcal{S}_{\mathcal{D}^{\star}}$ in order to respect the falsification principle, then we take the final average of elements in $\mathcal{S}_{\mathcal{D}^{\star}}$, which yields the spectra that are shown in this review as an example. To choose the algorithm parameters, we perform short preliminary runs during which we monitor the fitness as a function of the generation. A slow increase of the fitness usually indicates a poor choice of the frequency parametrization, while a behavior of the fitness showing frequent abrupt increases indicates a good choice of parameters. The typical resulting initial population size is $\mathcal{N}_{\overline{s}}\simeq 5000$, the typical number of resampled data is $\mathcal{N}_{d^\star}\simeq 200$, and the typical number of generations is $N_\text{gen}\simeq 10000$. Longer runs and population sizes are seldom needed, but are of course beneficial if larger computational power is available, provided overfitting ($|\Phi_{d^{\star}}(\overline{s})|\ll l+1$) is avoided. The typical number of frequency intervals depends on the desired resolution and quality of initial imaginary-time data, but it is of order $m\simeq 500$. We usually find that this is the most delicate parameter, but the choice of logarithmically spacing at high frequencies, as described in subsection \ref{subsec:frequency}, prevents the need for very large $m$.

We have performed several tests \cite{vitali_initio_2010} on exactly solvable analytical models suitably discretized
and ``dirtied'' with random noise to ``simulate'' real data.
Having in mind the $^4$He case, we have tried to reconstruct spectral functions
consisting of linear combinations of Gaussians, one sharp ``peak''
at small $\omega$ and one broad contribution at higher $\omega$, or, for the one--dimensional case, a combination of rectangular shapes and power-law decays \cite{bertaina_onedimensional_2016}. We have observed that none of the parameters have a critical role, once the frequency support is identified. However, only some features of the exact solution can be consistently reproduced: we have no possibility to exactly reconstruct the shape of $s(\omega)$, especially at high frequency;
on the other hand, access is granted to the 
identification of the presence of a sharp peak and to its
position, to the position of the broad contribution,
to some integral properties involving $s(\omega)$ and to its support. Moreover, we found that it is possible to estimate properties of the shape of the spectra (for example power-law decay close to thresholds \cite{bertaina_onedimensional_2016,motta_dynamical_2016}), once the low-frequency support is reliably estimated. In this case it is also crucial to analyze directly the elements in $\mathcal{S}_{\mathcal{D}^{\star}}$ and not only their final average.

\section{Applications: The dynamical structure factor of liquid $^4\mathrm{He}$}\label{sec:applications}
The GIFT method has been successfully applied to the study of spectral functions of various zero--temperature quantum systems in different geometries, such as $^4$He \cite{vitali_real_2009,vitali_initio_2010,minoguchi_accurate_2011,rossi_microscopic_2012,arrigoni_excitation_2013,bertaina_onedimensional_2016,motta_linear_2016}, $^4$He or H absorbed on various substrates \cite{reatto_novel_2012,nava_adsorption_2012,nava_superfluid_2012,reatto_novel_2013}, $^3$He \cite{nava_equation_2012,nava_dynamic_2013}, hard spheres \cite{rota_quantum_2013,rota_many_2014,motta_dynamical_2016}, soft particles \cite{saccani_bose_2011,saccani_excitation_2012,molinelli_roton_2016,teruzzi_microscopic_2016}, and the Fermi-Hubbard model \cite{vitali_computation_2016}. Moreover, a finite--temperature version of the GIFT method has been applied to the study of spectral functions for a system of $^4$He atoms in which Bose statistics has been suppressed \cite{minoguchi_density_2013}.

We present now some applications of this approach: for the sake of conciseness we will discuss only some of our results concerning the dynamical structure factor of $^4$He atoms in three-dimensions (3D) \cite{vitali_initio_2010} and confined to two-dimensional (2D) \cite{arrigoni_excitation_2013} or one-dimensional (1D) \cite{bertaina_onedimensional_2016} geometries. We mention that in the realm of bulk liquid helium, methods such as MaxEnt \cite{boninsegni_density_1996,baroni_reptation_1999}, modified kernels \cite{roggero_dynamical_2013}, and simulated annealing \cite{ferre_dynamic_2016}, have also been used. The dynamical structure factor $S(q,\omega)$ is a spectral function which is directly related to scattering experiments coupled to density fluctuations in linear response. It is a function of both frequency $\omega$ and momentum $q$ and its peaks allow for the determination of the dispersion relations of coherent collective modes, if present, while broad features indicate multiple excitations or damped modes.
The study of the spectrum of elementary excitations of $^4$He systems in different geometries
is of interest on one hand to investigate the fate of the phonon--maxon--roton spectrum 
upon change of the dimensionality of the system; on the other hand, 
it is useful for the interpretation of past or forthcoming scattering experiments 
involving $^4$He systems in bulk \cite{beauvois_superfluid_2016} or confined geometries: $^4$He atoms adsorbed on planar substrates \cite{lauter_mathrmhe_1992}
or confined in nano--channels \cite{savard_hydrodynamics_2011,taniguchi_dynamical_2013}.

In all the applications we are going to discuss in the following, the intermediate scattering functions $F(q,\tau)$,
i.e. the basic ingredient of the GIFT method,
have been computed using the exact shadow path integral ground state (SPIGS) method \cite{galli_recent_2003,galli_shadow_2004}.
For the technical details we invite the reader to refer to 
the original articles \cite{vitali_initio_2010,arrigoni_excitation_2013,bertaina_onedimensional_2016}.
In short, SPIGS is a $T=0$~K path integral ground state (PIGS) method \cite{sarsa_path_2000}, which relies on an imaginary-time projected Shadow wave function \cite{vitiello_variational_1988}.
Path integral projector methods like PIGS and SPIGS allow for the calculation of exact ground-state expectation 
values by systematically improving an approximation of the ground-state wave function via successive 
small imaginary-time projections.
The imaginary-time evolution cleans up the spurious overlaps of the variational wave function 
with excited states and introduces the missing ground-state correlations between particles.
We have shown that when the total imaginary-time projection is large enough, 
these methods provide exact ground-state expectation values, within the statistical uncertainties of the calculations,
without any bias due to the choice of the variational wave function \cite{rossi_exact_2009}.

\subsection{Three-dimensional case}\label{subsec:3d}
The first application of the GIFT method was the study of the dynamical structure factor
of superfluid $^4$He in 3D geometry \cite{vitali_initio_2010}.
As far as we know, this was the first analytic continuation method, different from the MaxEnt method,
applied to this very peculiar condensed matter system. The use of GIFT turned out to be 
a major improvement with respect to previous MaxEnt studies of superfluid $^4$He \cite{boninsegni_density_1996}:
we were able to recover sharp quasi--particle/elementary excitations, with excitation energies 
in good agreement with experimental data, and spectral functions displaying also the multi--phonon branch 
(i.e. a branch of the spectrum corresponding to the creation of multiple elementary excitations) 
with the correct relative spectral weight.

\begin{figure*}[t]
 \centering
 \subfigure{\centering
 \includegraphics[width=\figwid]{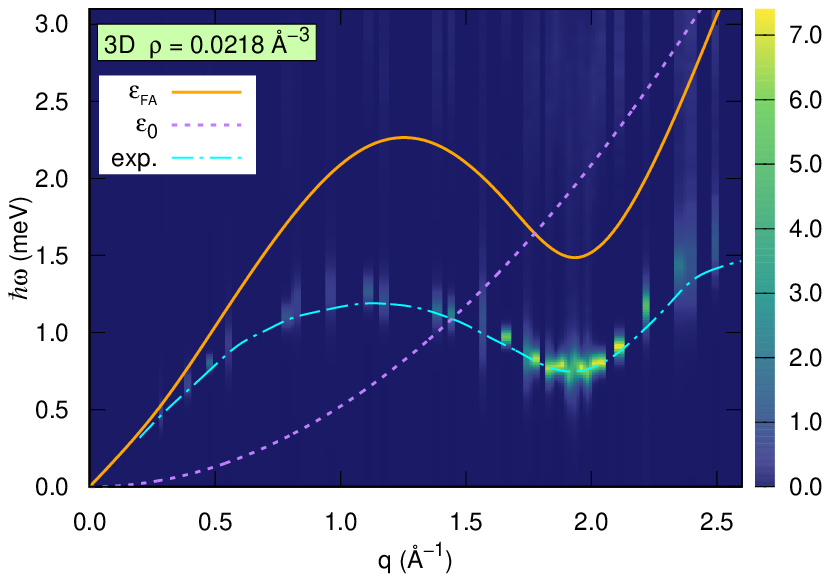}
 }
 ~
 \subfigure{\centering
 \includegraphics[width=\figwid]{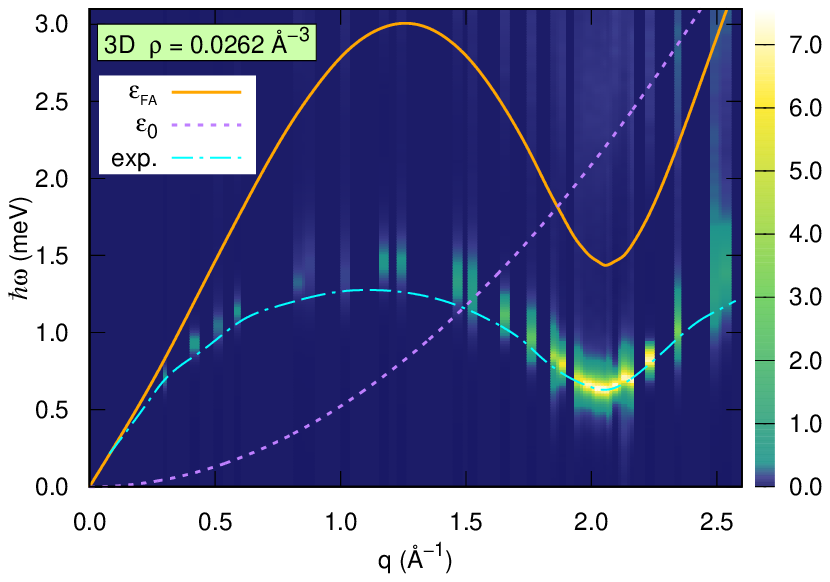}
 }
 \\
 \subfigure{\centering
 \includegraphics[width=\figwid]{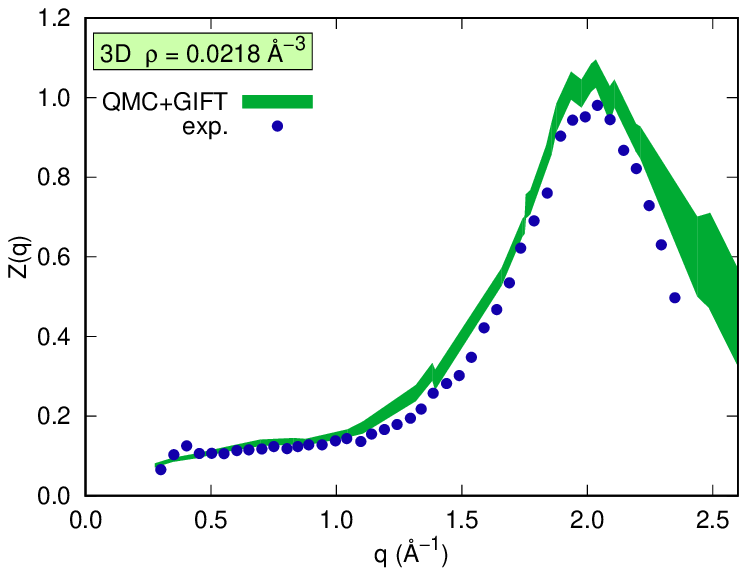}
 }
 ~
 \subfigure{\centering
 \includegraphics[width=\figwid]{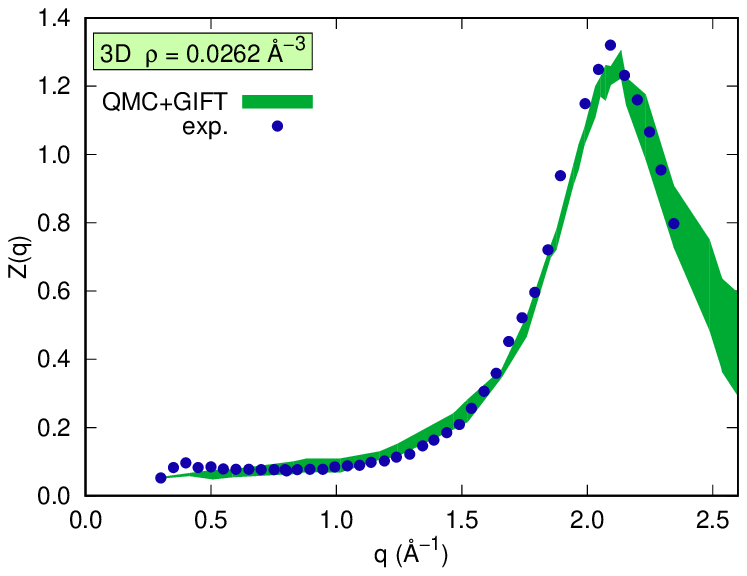}
 }
 \caption{\label{fig:spectra3D}
{\bf Dynamical structure factor of zero--temperature superfluid $^4$He in 3D geometry}.
(Upper panels) $S(q,\omega)$ extracted at the equilibrium $\rho_{\mathrm{3D}}=0.0218$ \AA$^{-3}$ and freezing $\rho=0.0262$ \AA$^{-3}$ bulk densities (from \cite{vitali_initio_2010}),
for a discrete set of wave vectors compatible with the periodic boundary conditions used in the simulations.
Color scale represents bins' height in units of $\hbar/$meV. Weights exceeding the color scale are cropped.
(Lower panels) Corresponding GIFT strength of the quasi--particle peak $Z(q)$ as a function of $q$ and comparison to
experimental data \cite{gibbs_collective_1999,cowley_1971,woods_1973}. The reference free $\varepsilon_0(q)$ and Feynman's $\varepsilon_{\mathrm{FA}}(q)$ dispersion relations are described in the text.}
\end{figure*}

These results can be observed in Fig.~\ref{fig:spectra3D} where color maps of the 
dynamical structure factors at equilibrium ($\rho_{\mathrm{3D}}=0.0218$ \AA$^{-3}$) and freezing
($\rho_{\mathrm{3D}}=0.0262$ \AA$^{-3}$) densities, extracted with the GIFT method, are shown together with the experimental
quasi--particle/elementary excitations energies \cite{gibbs_collective_1999,cowley_1971,woods_1973}. As a reference, we also plot the free--particle
dispersion $\varepsilon_0(q)=\hbar^2 q^2/2m$, where $m$ is $^4$He mass. Moreover, we show the famous Feynman's dispersion relation $\varepsilon_{\mathrm{FA}}(q)=\varepsilon_0(q)/S(q)$, which can be derived using a variational argument \cite{feynman_atomic_1954}, where $S(q)$ is the static structure factor. This dispersion correctly manifests a minimum in the excitation close to momentum $2\pi/a$, where $a$ is the hard-core size of the interaction potential, and was first phenomenologically hypothesized by Landau \cite{landau_theory_1941}. In the same momentum region, the static structure factor features a peak, provided the average interparticle distance is of the order of $a$. 
Our reconstructed $S(q,\omega)$ exhibit an overall structure
in good agreement with experimental data: a sharp quasi--particle
peak and a shallow multi--phonon maximum are present.
Both features appeared for the first time within an analytic continuation procedure applied to a QMC study
of a many--body system in the continuum.

By integrating the extracted $S(q,\omega)$, one has access to quantities like the
strength of the single quasi--particle peak, $Z(q)$, and thus also to the contribution to
the static structure factor, $S(q)$, coming from multi--phonon excitations.
Remarkably, $Z(q)$ is in close
agreement with experimental data \cite{gibbs_collective_1999} (see Fig.~\ref{fig:spectra3D}), thus
strongly suggesting that the broad structure in $S(q,\omega)$ at large frequency
carries indeed reliable physical information on the multi--phonon branch of the spectrum.
Given the assumption of a pair-wise interatomic interaction \cite{aziz_accurate_1979} and the 
experimental and algorithmic statistical uncertainties (the latter being estimated via multiple
independent QMC simulations and GIFT reconstructions, also involving variants of the interaction potential \cite{aziz_accurate_1979,aziz_new_1987,aziz_initio_1995}), the agreement of the extracted $Z(q)$
is very good. This shows that, via analytic continuation, it is at least possible to extract one sharp feature in the spectral function with the correct spectral weight and a broad multi--phonon component
which represents semi--quantitatively the combination of multiple quasi--particle excitations.
Note that here the width of the reconstructed single quasi--particle peak
is mainly a measure of the uncertainty in the
statistical reconstruction of the position. Thus the exact shape of the spectral function is not accessible, given the ill-posed nature of the problem: future improvements will unavoidably require QMC simulations on more powerful computational facilities.

\subsection{Two-dimensional case}\label{subsec:2d}
Another application of the GIFT method to superfluid $^4$He systems has been the study
of $S(q,\omega)$ for a pure 2D geometry.
As highlighted by the 2016 Nobel Prize in Physics \cite{Nobel_physics_2016},
bosons in two dimensions are of great theoretical interest because the standard scenario of superfluidity associated
with Bose-Einstein condensation (BEC) is not appropriate anymore.
As shown by J.M. Kosterlitz and D.J. Thouless \cite{kosterlitz_ordering_1973},
the notion of long-range order has to be replaced by
that of topological long-range order, characterized by a slow algebraic decay of the local order parameter correlation function.
Notwithstanding a vanishing order parameter in 2D, i.e. a condensate wave function which 
vanishes at any finite temperature for a bulk system, a superfluid response is theoretically 
predicted up to a temperature where vortex and antivortex pairs unbind.

\begin{figure*}[t]
 \centering
 \subfigure{\centering
 \includegraphics[width=\figwid]{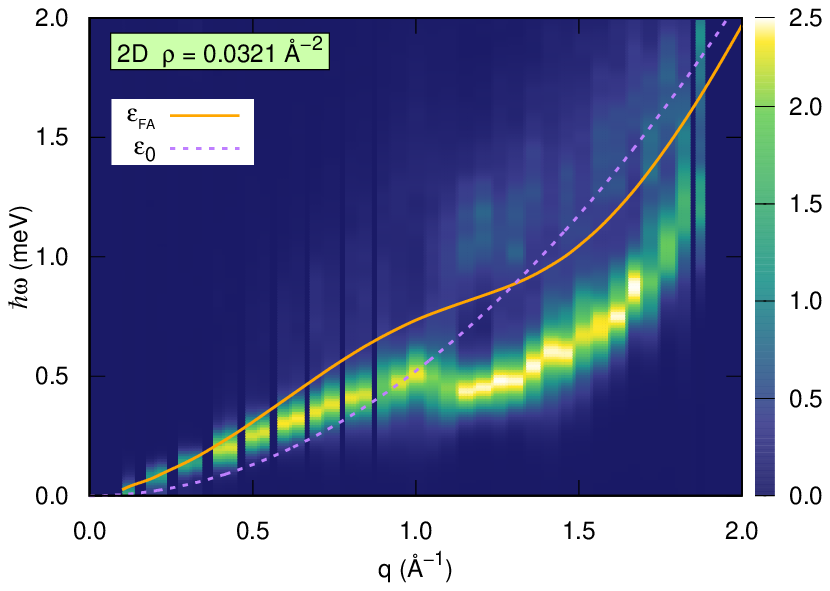}
 }
 ~
 \subfigure{\centering
 \includegraphics[width=\figwid]{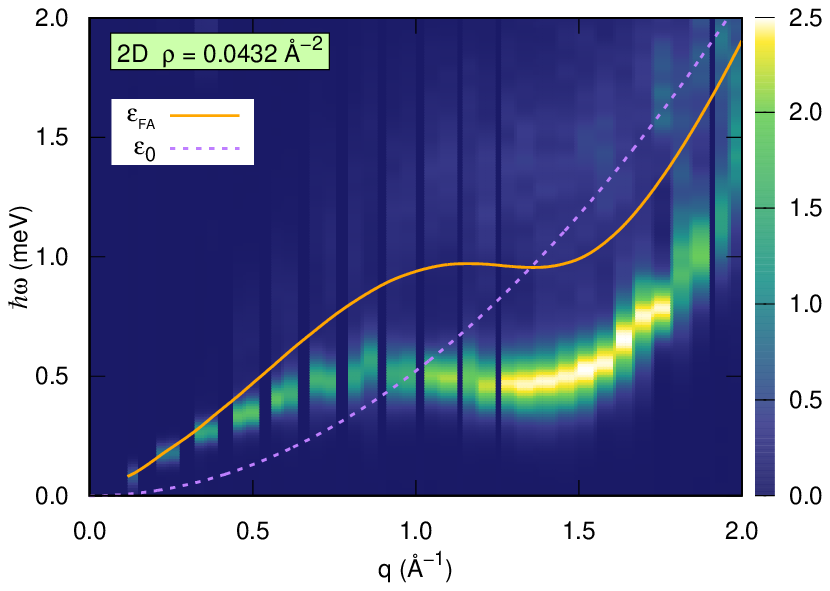}
 }
 \\
 \subfigure{\centering
 \includegraphics[width=\figwid]{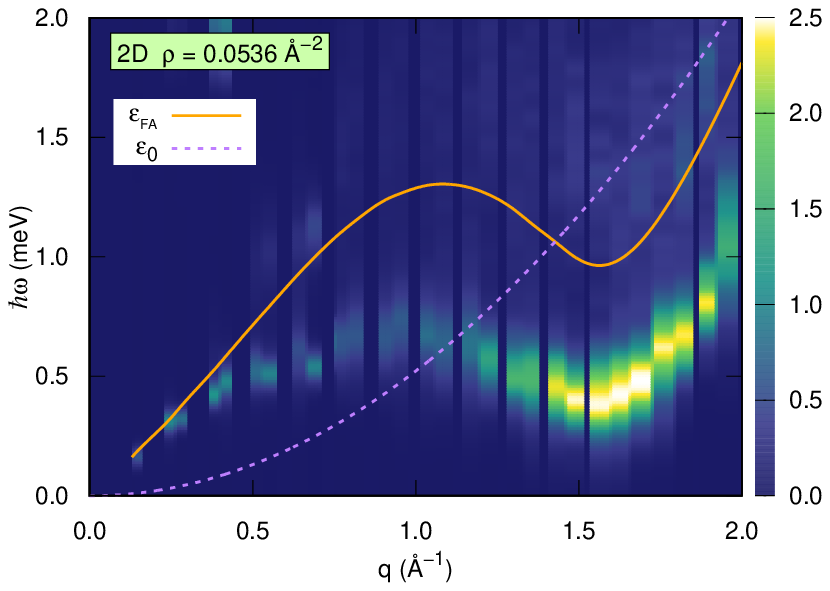}
 }
 ~
 \subfigure{\centering
 \includegraphics[width=\figwid]{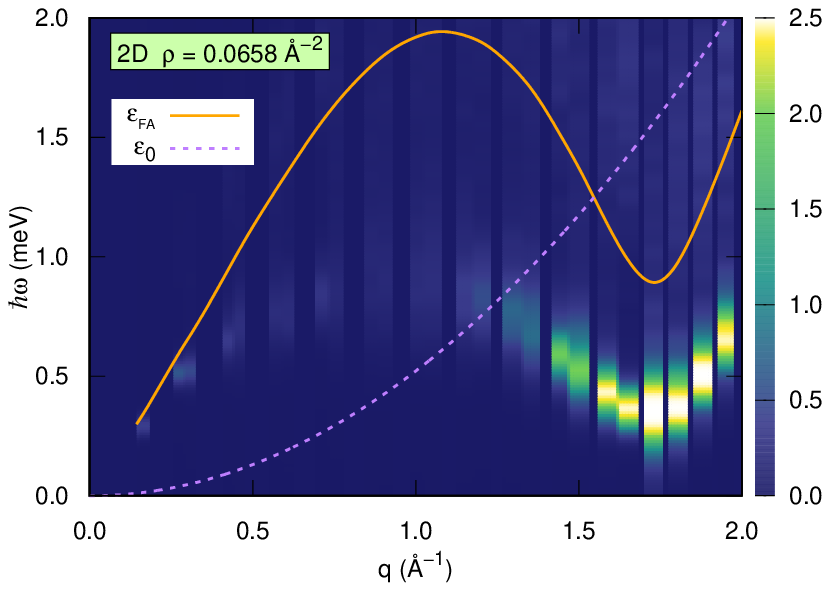}
 }
 \caption{\label{fig:spectra2D}
{\bf Dynamical structure factor of zero--temperature superfluid $^4$He in 2D geometry}.
$S(q,\omega)$ extracted at the areal densities $\rho_{\mathrm{2D}}=0.0321$ \AA$^{-2}, 0.0432$ \AA$^{-2}, 0.0536$ \AA$^{-2}, 0.0658$ \AA$^{-2}$ (from \cite{arrigoni_excitation_2013}). See also caption of Fig.~\ref{fig:spectra3D}.
}
\end{figure*}
The dynamical structure factors obtained for 2D $^4$He \cite{arrigoni_excitation_2013} at
different areal densities are reported in Fig.~\ref{fig:spectra2D}.
We found well defined excitations in the full density range where the system is superfluid;
however, significant differences are present with respect to 3D $^4$He.
In 2D, close to the spinodal density ($\rho_{\mathrm{2D}}=0.0321$ \AA$^{-2}$), where the system is unstable against droplet formation, the excitation spectrum features the maxon (the maximum of the coherent dispersion) and the roton (the finite-momentum minimum) frequencies almost coalescing in a plateau. At the equilibrium density ($\rho_{\mathrm{2D}}=0.043$ \AA$^{-2}$), the small peak in the static structure factor causes a maxon--roton structure which is rather
weak, with the maxon frequency only 10\% higher than the roton frequency. Above the equilibrium density, a well defined maxon--roton structure develops
(see the density $\rho_{\mathrm{2D}}=0.0536$ \AA$^{-2}$)
and, finally, at freezing density ($\rho_{\mathrm{2D}}=0.0658$ \AA$^{-2}$)
the ratio between maxon and roton energies is found as large as 3.
At the same time the wave vector of the roton has a strong density
dependence, whereas that of the maxon is almost density independent. 
This strong evolution of the shape of the excitation spectrum with the density is probably 
due to the wider density--range of existence of the fluid phase in 2D:
the freezing density is more than twice the spinodal density while in 3D it is only 60\% larger.
Moreover, in the maxon region for densities above equilibrium,
the quasi--particle excitation peak is substantially broadened with respect to the roton region. This implies that, over an extended region of wave--vectors and of density, the elementary excitations have a finite lifetime even at $T=0$~K. In fact, they can decay into other excitations, since the phonon region is characterized by a strong anomalous dispersion, featuring a positive curvature \cite{arrigoni_excitation_2013}.

\begin{figure*}[t]
 \centering
 \subfigure{\centering
 \includegraphics[width=\figwid]{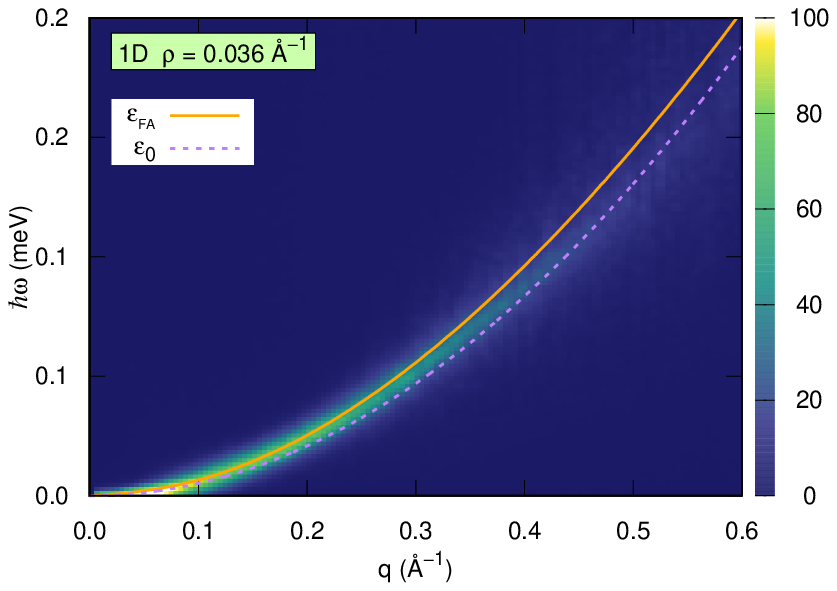}
 }
 ~
 \subfigure{\centering 
 \includegraphics[width=\figwid]{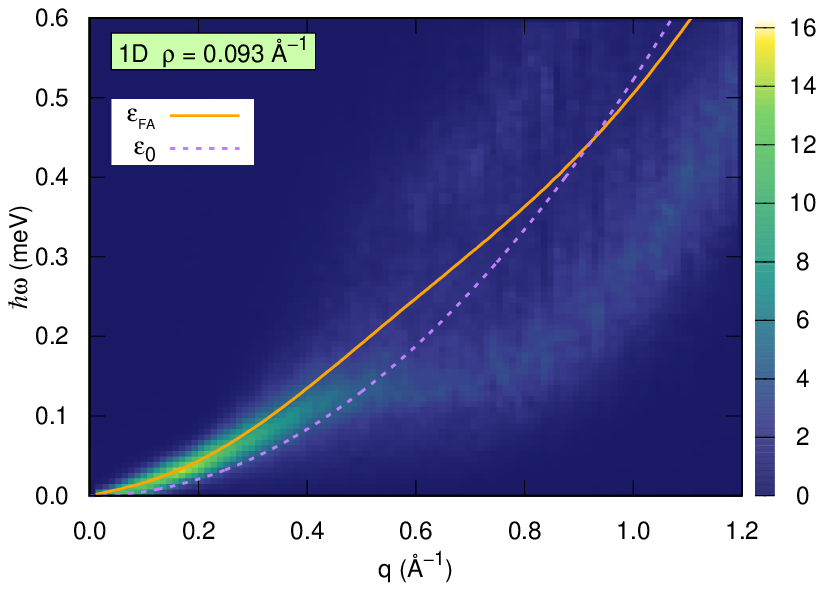}
 }
 \\
 \subfigure{\centering 
 \includegraphics[width=\figwid]{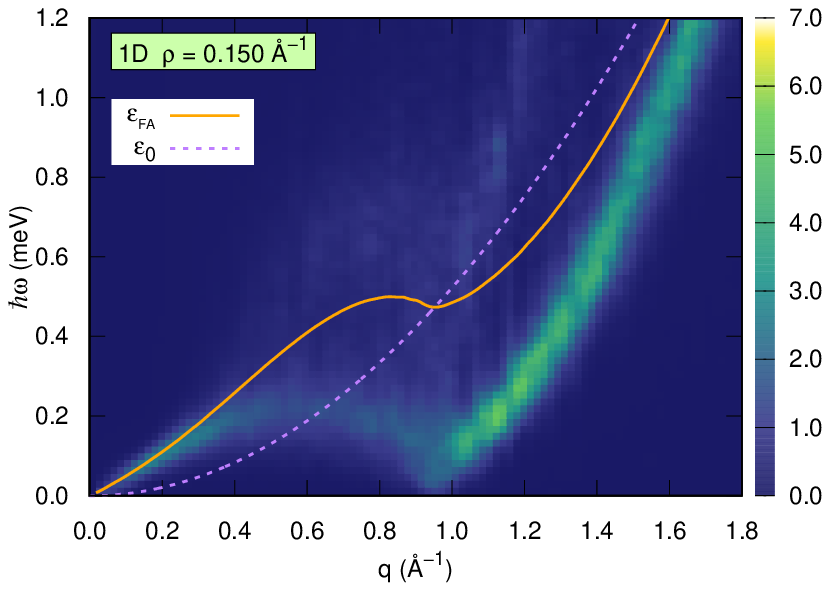}
 }
 ~
 \subfigure{\centering
 \includegraphics[width=\figwid]{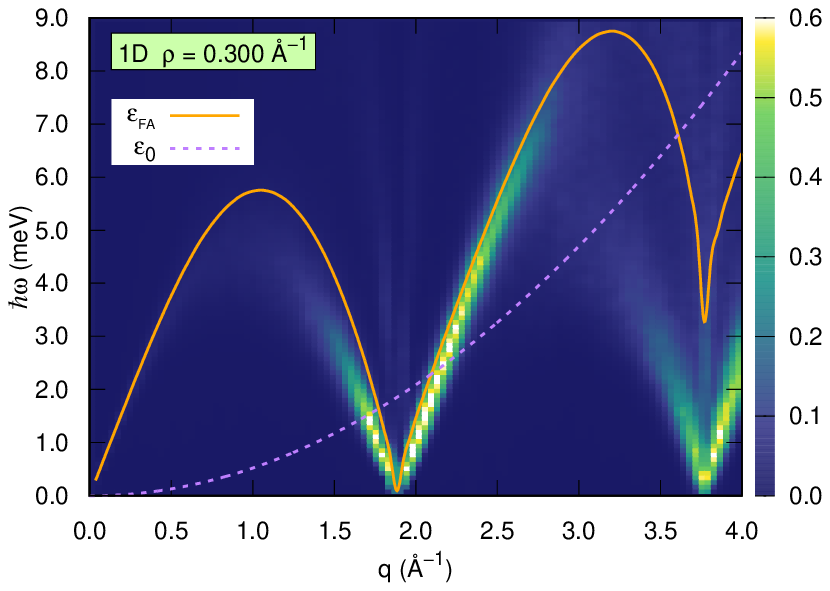}
 }
 \caption{\label{fig:spectra1D}
{\bf Dynamical structure factor of zero--temperature superfluid $^4$He in 1D geometry}.
$S(q,\omega)$ extracted at the linear densities $\rho_{\mathrm{1D}}=0.036$ \AA$^{-1}, 0.093$ \AA$^{-1}, 0.150$ \AA$^{-1}, 0.300$ \AA$^{-1}$ (from \cite{bertaina_onedimensional_2016}). See also caption of Fig.~\ref{fig:spectra3D}.
}
\end{figure*}

\subsection{One-dimensional case}\label{subsec:1d}
To conclude our discussion on the applications of the GIFT method to $^4$He systems,
we briefly review our recent study on a system of $^4$He atoms 
in a pure 1D geometry.
One-dimensional quantum systems exhibit spectacular signatures of the interplay between quantum fluctuations,
interaction and geometry. The reduced dimensionality prevents the spontaneous breaking of continuous symmetries in the presence of short-range interactions \cite{mermin_absence_1966}, which results in a single Luttinger liquid phase for $^4$He, with different character depending on density. Moreover, in the presence of hard--core interactions, bosonic and fermionic systems start to share common behavior, since two-body correlations must decay to zero in both cases.
A color map of $S(q,\omega)$ for the 1D system is shown in Fig.~\ref{fig:spectra1D}.
By increasing the density, the dynamical structure factor reveals a transition from a 
highly compressible liquid near the equilibrium linear density ($\rho_{\mathrm{1D}}=0.036$ \AA$^{-2}$) to a quasi--solid regime
($\rho_{\mathrm{1D}}=0.300$ \AA$^{-2}$). Notice that the range of considered densities is much larger than in higher dimensions, so the typical frequency and momentum scales vary considerably.

In the low-frequency limit, the dynamical structure factor can be described by the quantum
hydrodynamic Luttinger-liquid theory \cite{giamarchi_quantum_2003}: elementary excitations are unavoidably collective, i.e. phonons (this holds true even for fermionic systems).
At higher energies, the GIFT analytic continuation approach 
provides quantitative results beyond Luttinger-liquid theory.
In particular, as the density increases, the interplay between dimensionality
and interaction makes $S(q,\omega)$ manifest a pseudo-particle-hole continuum typical
of fermionic systems.
The fate of the phonon--maxon--roton spectrum, which characterizes the excitations in higher dimensions,
is to merge into a pseudo-particle-hole continuum.
It is interesting to note that, by increasing density, the spectral weight moves towards lower frequencies for wave--lengths of the order of the average interparticle distance, similarly to the behavior in higher dimensions. However, instead of having a neat roton excitation, a broad spectral structure 
bends down, and only at very high linear densities almost coherent modes are reached. However, we mention in passing that a power-law behavior close to the lower spectral support is in fact observed and indeed expected from non-linear Luttinger theories: we refer the interested reader to Refs.~\cite{bertaina_onedimensional_2016,motta_dynamical_2016}, where analytical efforts, motivated by the obtained GIFT spectra, yielded remarkable results.

\section{Conclusions}
We have described in detail a strategy we have developed to face the
analytic continuation problem that emerges whenever real-time dynamics
of strongly correlated physical systems is studied relying on 
estimations of imaginary-time correlation functions.
We have presented it in a pedagogical way, in order to
allow researchers to take full advantage of the methodology.
We have enriched the presentation with figures
showing the very accurate results we have obtained for systems
of $^4$He atoms in different geometries. In general, the family of stochastic analytic continuation methods is becoming more affordable due to the increase of available computational power, and is thus expected to have higher impact in the future, due to its accuracy. Our method combines the genetic speed-up coming from the {\it Crossover} move, to a remarkable robustness with respect to the chosen parameters, coming from the averaging procedure and the {\it Rejection} step. The only crucial input to be optimized is the frequency support and parametrization, which suggests that further improvement of the method is foreseeable. Moreover, knowledge about the typical features of the considered spectra helps in choosing the rate at which appropriate operators are preferentially called, thus increasing the rate of convergence.
The problem we have faced belongs to 
the huge class of inverse problems, a deep topic also from a fundamental epistemological point of view \cite{popper2002logic}. The basic idea of the falsification principle \cite{tarantola_popper_2006} guided us in 
our particular implementation of analytic continuation: moreover, every analytic continuation problem emerging in Physics, applied Mathematics or, more generally, applied Science, can in principle be tackled by a suitable 
variation of the GIFT algorithm, which is \emph{per se} an approach suitable for hybridization with other methods. We thus expect that the key ideas underlying our approach can be efficiently used also in other inverse/analytic continuation problems and in many research fields, like, just to cite one example, image reconstruction.
In fact, one of us has recently faced the Phase Problem in Coherent Diffractive Imaging, following a similar approach:
we have found that by building a memetic algorithm, i.e. hybridizing a Genetic Algorithm 
with standard iterative methods, it is possible to outperform the phase retrieval capabilities 
of the algorithms used as \emph{memes} to assist the genetic stochastic search~\cite{MPR_article_2016}.

\subsection{Acknowledgements}
We thank our collaborators L. Reatto, M. Rossi, M. Nava, F. Tramonto and M. Motta, for discussions during the development of the method. We acknowledge the CINECA award IsC29-SOFTDYN and the CINECA and Regione Lombardia LISA 
award LI05p--PUMAS, for the availability of high--performance computing resources and support.

\subsection{Funding}
This work was supported by the NOXSS PRIN (2012Z3N9R9) project of MIUR, Italy.
We acknowledge support from the University of Milan through Grant No. 620, Linea 2 (2015). E.V. acknowledges support from the Simons Foundation and NSF (Grant no. DMR-1409510).
G.B. also acknowledges support by ECT$^*$ during the workshop ``Advances in transport and response properties of strongly interacting systems'' (Trento, Italy, 2016).


\end{document}